\documentclass[manuscript]{aastex}
\usepackage{epstopdf}
\usepackage{graphics}
\usepackage{float}
\usepackage{hyperref}

\shorttitle{A passive FPAA based RF scatter meteor detector}
\shortauthors{Djorgovski et al.}

\begin{document}

\title{A passive FPAA based RF scatter meteor detector}

\author{A. Popowicz}
\affil{Silesian University of Technology, Institute of Automatic Control, Poland, Gliwice, Akademicka 16}
\email{apopowicz@polsl.pl}

\and
\author{A. Malcher, K. Bernacki}
\affil{Silesian University of Technology, Institute of Electronics, Poland, Gliwice, Akademicka 16}
\email{amalcher@polsl.pl, kbernacki@polsl.pl}

\and

\author{K. Fietkiewicz}
\affil{Polish Fireball Network}
\email{parmo.pfn@gmail.com}

\begin{abstract}
In the article we present a hardware meteor detector. The detection principle is based on the electromagnetic wave reflection from the ionized meteor trail in the atmosphere. The detector uses the ANADIGM field programmable analogue array (FPAA), which is an attractive alternative for a typically used detecting equipment - a PC computer with dedicated software. We implement an analog signal path using most of available FPAA resources to obtain precise audio signal detection. Our new detector was verified in collaboration with the Polish Fireball Network - the organization which monitors meteor activity in Poland. When compared with currently used signal processing PC software employing real radio meteor scatter signals, our low-cost detector proved to be more precise and reliable. Due to its cost and efficiency superiority over the current solution, the presented module is going to be implemented in the planned distributed detectors system. 
\end{abstract}


\section{Introduction}

\begin{figure}[H]
\centering
\epsscale{0.75}
\plotone{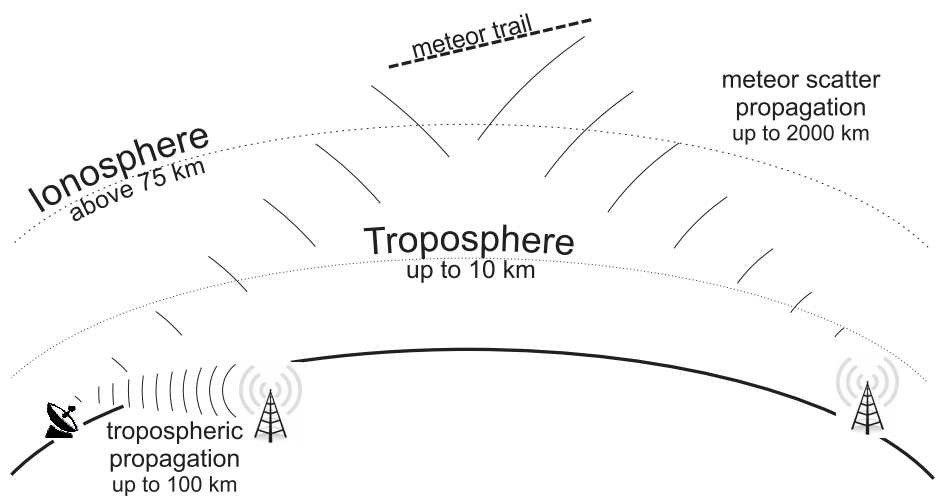}
\caption{The principle of meteor scatter phenomenon.}
\label{f1}
\end{figure}

One of the methods used for meteor detection is the radio meteor observation \cite{mck61,wei88,wis95}. The principle is based on the electromagnetic waves reflection from the meteor's trail in the atmosphere, which is called the meteor scattering. When a meteor enters the atmosphere, it ionizes the electrons in atmospheric E-Layer (90-150km, Ionosphere layer) and leaves a plasma trail which, for a short time, works like a mirror for radio waves. Although the effect of the scattering is the strongest for 20-50 Mhz frequencies, it is possible to receive reflected signals from distant media radio stations operating in higher frequency range (90Mhz and above). This makes the phenomenon observable even by amateurs with a relatively low-cost equipment. The phenomenon of scattering is utilized extensively for meteor burst communication, which is a radio propagation mode allowing to establish short communication windows between distant stations. While the service of local media stations (like TV or FM) extends only slightly the optical horizon, reaching tens of kilometers, the maximum range the meteor scattering allows reception of signals from stations up to 2000km away \cite{Sugar}. The phenomenon is visualized in Fig. \ref{f1}.

\begin{figure}[H]
\centering
\epsscale{0.7}
\plotone{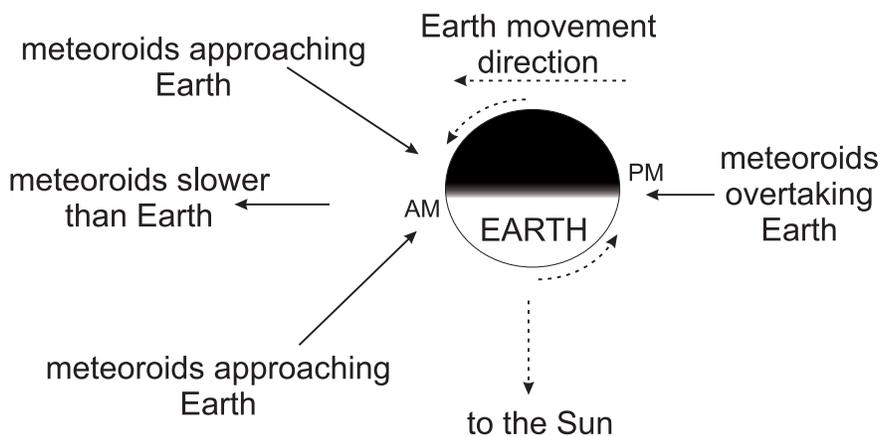}
\caption{The visualization of the Earth in meteoroids environment.}
\label{f2}
\end{figure}

The number of meteoroids entering the atmosphere per hour changes periodically during the day.  While in the morning we can observe the meteors approaching the Earth and the ones which are slower than our planet, in the evening only the meteoroids faster than the Earth enter the atmosphere. This fact causes characteristic diurnal variations of the meteor rate. The principle of this phenomenon is depicted in Fig. \ref{f2}.

In contrast to visual techniques, like video registration, which can not be carried out during the day, full moon or cloudy weather, the radio observations of meteors are of great interest as they can be performed continuously. The obtained count histogram is scientifically valuable and provides a number of meteor characteristics which require the analysis of the underdense and the overdense meteor distributions \cite{Rodriguez}. We would like to note that the overdense and the underdense meteors are the results of respectively heavy and light particles entering the atmosphere.  The main coefficient - the mass index - provides information about the composition of meteoroids and their origins. While the high index is observed for showers from remnants of comets, the low index suggests the planetoid origins. The brief review of tasks, which may be accomplished employing the count histogram, is given below.
 
\begin{figure}[H]
\centering
\epsscale{1}
\plotone{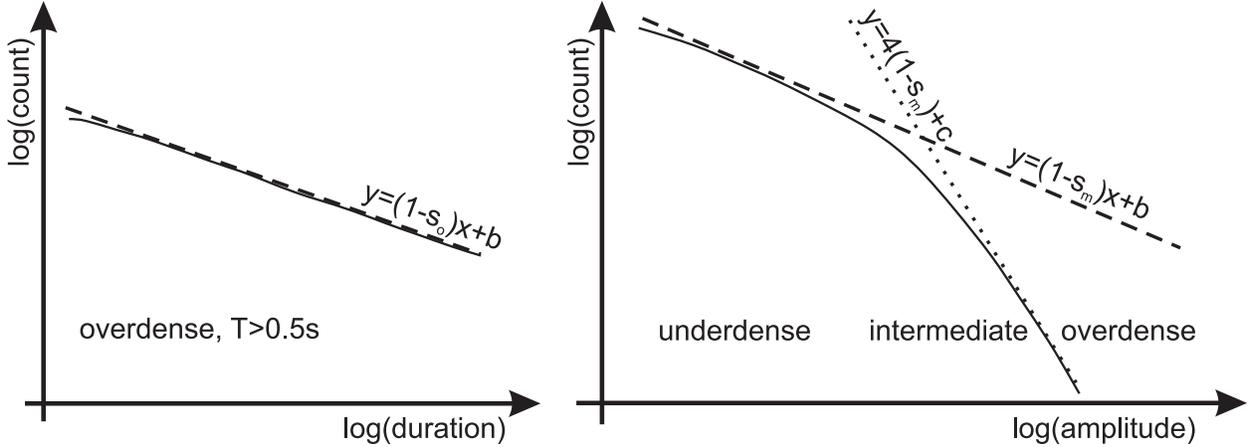}
\caption{Mass index calculations. On the left - the overdense mass index estimation utilizing phenomena duration, on the right - the mean mass index estimation from amplitude histogram.}
\label{lf3}
\end{figure}

 \begin{enumerate}
 \item The time occurrences of maximum of meteor showers are compared with the predictions to obtain the time-shifts whose origins have to be investigated.
 \item New meteor showers, observed in the form of abnormal activity, may be discovered.
 \item The overdense meteor mass distribution may be obtained by plotting logarithm T (the duration of phenomena) against the logarithm of the number of meteors with duration exceeding T. Only the phenomena longer than 0.5 seconds have to be included here. The slope should be $1-s_o$, where $s_o$ is the mass index of overdense meteors (see the left plot in Fig \ref{lf3}). 
 \item The mean mass index, which includes information from small and large particles, may be derived from the dependency between the logarithm of phenomenon amplitude and the number of detections above the given amplitude. The method of calculation of mean mass index $s_m$ is explained graphically on the right plot of Fig. \ref{lf3}. One can obtain required data in two ways: directly, by investigating detected signal amplitude, or indirectly, by utilizing parallel detectors set to various sensitivities.
 \end{enumerate}

It has to be mentioned that by using a more sophisticated transmitter - receiver pair (active radar) some further properties of the meteoroids can be obtained. The array of receivers and high-speed signal correlation computations can provide information about meteoroid velocity vector \cite{ker12}. Moreover, some observations of the ablation and the fragmentation of the meteoroids are possible with a specially designed radar system \cite{mal11}.

\begin{figure}[H]
\centering
\epsscale{0.7}
\plotone{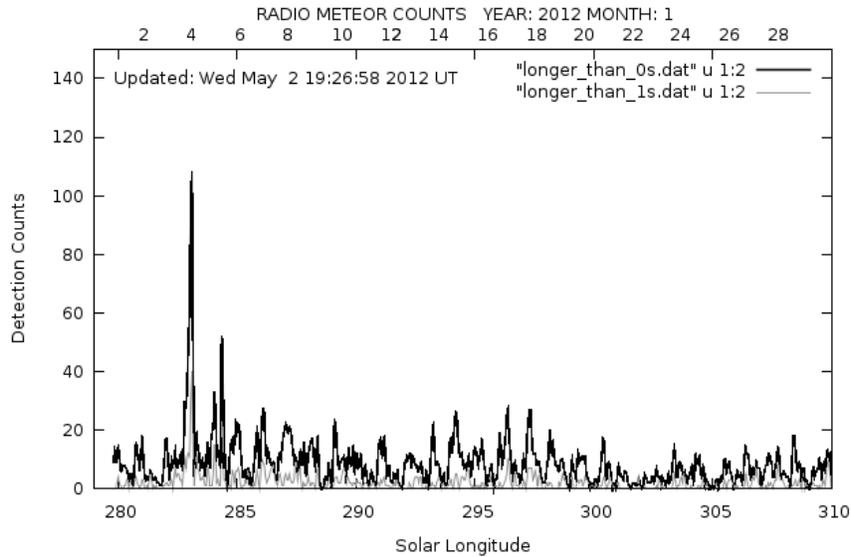}
\caption{Monthly detection histogram from the Polish Fireball Network (January 2012). Light gray line - detections longer than 1 s, black line - detections longer than 0 s (all detections) \cite{pfn13}.The peak refers to the Quadrantids meteor shower on 3$^{rd}$ January 2012. The measurements were binned in 0.1 solar longitude time intervals (about 2.43 hour).}
\label{f4}
\end{figure}

In our project, we concentrate on a passive radio detection circuit for Polish Fireball Network (PFN) research activities. The PFN is the organization which continuously monitors meteor showers and fireballs over Poland \cite{PFN}. The network consists of several base stations located all over the country, equipped with high-speed camera systems, recording and analyzing images in real time. They also use passive radio detection methods to enable observations during bad weather conditions or during the day \cite{Zloczewski, Fietkiewicz}. In Fig. \ref{f4} we show an exemplary monthly radio count histogram provided by PFN, which includes Quadrantids meteor shower. For this purpose they used a PC computer, specially developed software (MetAN) and a Realistic DX-394 receiver tuned to distant, east European radio stations. 

The PFN PC computers running 24 hours a day and detecting meteor radio signals are neither energy efficient nor portable solutions. It is to be noted that while a single station is still able to provide useful information about meteors, a system of multiple detectors, located at different sites and directed to various distant stations, may significantly improve the measurement efficiency and extend the analysis capabilities. These two main issues are discussed in detail below.

\begin{enumerate}
\item Firstly, we should note that the detector robustness depends on local electromagnetic environment. The sites far from strong radio stations are generally preferred since there are fewer false detections resulting from interfering frequencies. However, the best sites usually suffer from lack of access to the power, thus ideally the detection module should be self-sufficient e.g. powered by solar cells. 
\item The reception of reflected signal depends on the relative position of the receiver, meteor trail and remote radio station. By directing the detector toward different stations we would gain additional information about the brief position of trail relatively to the detector. Moreover, a distributed system of detectors, spread over hundreds of kilometers, should provide far better meteor localization capabilities and would work like a radar. We plan to investigate this issue in the future.
\end{enumerate}

In the article we introduce a dedicated detection electronic circuit, easy connectible to any radio receiver, which can work as a module in a remote, battery powered, meteor scatter detector. For our task we selected a field programmable analog array (FPAA) manufactured by ANADIGM. We also intend to show that for such a purpose an analogue array is an interesting alternative to commonly used digital signal processing units.

The circuit is dedicated to planned distributed system of remote meteor detectors. We also consider the amateur application of our module. With a growing number of astro-amateurs and dedicated self-sufficient equipment, it may be possible to explore the meteor phenomena as it is performed for variable stars in the American Association of Variable Star Observers (AAVSO) \cite{AAVSO}.

The article is organized as follows. In Chapter 2 we introduce the internal structure of the ANADIGM FPAA circuit and its principle of operation. Our signal chain implemented for meteor detection is explained in detail in Chapter 3. Finally, in Chapter 4, we describe our verification methodology based on the comparison with the results obtained by MetAn software, which is currently used for meteor scatter detection by the Polish Fireball Network. The results of the tests are presented and analyzed in Chapter5. We summarize the results of our research in Chapter 6.

\section{ANADIGM Field Programmable Analog Array}

The AN231E04 device represents the third generation of the dynamically reconfigurable FPAA AnadigmApex offered by Anadigm® \cite{res1}. It is powered by 3.3 V supply voltage.  The previous FPAA family, called AnadigmVortex, was powered by 5V, but its DC performance was much worse.

The FPAA circuit consists of four configurable analog blocks (CAB) built on the switched-capacitance principle, surrounded by a network of programmable connections. There are also seven programmable input/output cells on the silicon chip. Three of them can have access to a specialized chopper amplifier resource, which allows accurate amplification of very low energy input signals. All internal signal paths of the block are differential. 

IO cells contain both passive and active circuitry, which allows direct signal input and output, building active filters, sample and hold circuits, digital inputs, and digital outputs. The frequency response of continuous time input and output filters is determined by a combination of internal programming and external components.

\begin{figure}[H]
\centering
\epsscale{0.7}
\plotone{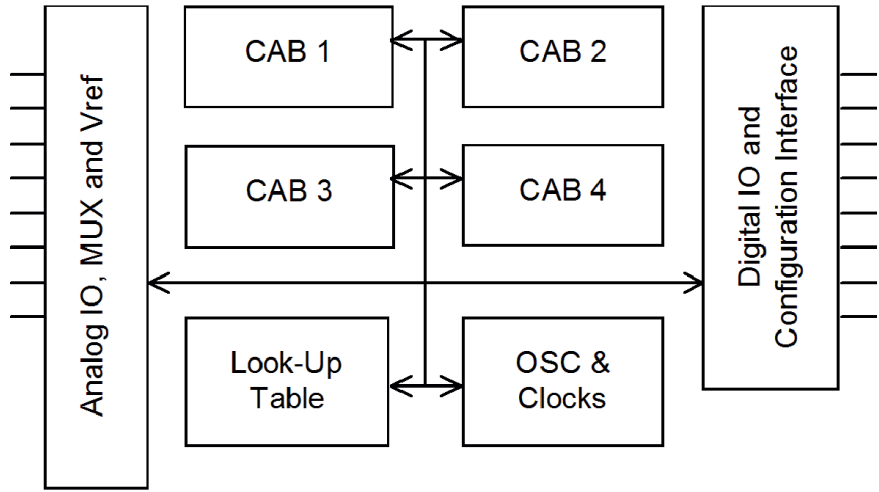}
\caption{Internal structure of ANADIGM field programmable analogue array.}
\label{f5}
\end{figure}

The block diagram of the AN231E04 device is shown in Fig.~\ref{f5}, where one can see its basic elements: configurable analog blocks (CAB), input and output cells, look up table (LUT), clocking circuits, communication interfaces and reference voltage sources.

The new AN231E04 devices can be reconfigured dynamically (on the run), which enables e.g. auto ranging, auto calibration and multimode work (allows to build many functions in a single IC). The microcontroller connected to the AN231E04 device can load the new device configuration data to the device still working in its previous configuration. The configuration data can then be swept in a single clock cycle.  This new and prospective feature of the dynamic reconfiguration ability of the AN231E04 devices enables the creation of a modern analog system, which can be (completely or in part) rebuilt in real time.

We chose this particular FPAA because of its proved very good capability of audio signal processing \cite{fal11,fal10}. By using different configurations and changing individual parameters we were able to create a circuit well suited for our purpose. Such designing flexibility makes FPAA a very attractive prototyping device. Moreover, the FPAA devices have been recently used as hardware detectors of QRS complexes in the ECG signal \cite{mal10,mal09}, which is an application similar to our audio meteor detector.

In comparison with popular Field Programmable Gate Arrays (FPGA), which also might be utilized as a meteor detector, FPAAs have several advantages. Below, we enumerate the most crucial ones for our project.

\begin{enumerate}
 \item FPAA devices introduce much lower electromagnetic disturbance, which should be considered when analyzing weak analogue signals \cite{Mocha}.
 \item The implementation process is very simple and efficient as the construction of signal path may be accomplished with an easy-to-use, drag-and-drop, diagram based environment - \emph{AnadigmDesigner} (FPGAs have to be programmed using dedicated VHDL or Verilog register transfer languages).
 \item Nearly all modifications of the existing signal chain may be tested by changing single values in user-friendly environment (the properties are clear and simple, like filter bandwidth or gain).
 \item The analogue filters implemented in FPAA are more reliable and efficient in contrast to their digital versions (the infinite response IIR digital filters suffer from low stability while the finite response FIR implementations, due to the high number of parallel calculations, usually require specialized high-efficiency units, like digital signal processors DSP ).
 \item The FPAAs do not require any analogue-to-digital conversions since they operate in the analogue domain, therefore the number of elements and the total cost of the circuit is lower.
\end{enumerate}

\section{Detection circuit}

\begin{figure}[H]
\centering
\epsscale{0.65}
\plotone{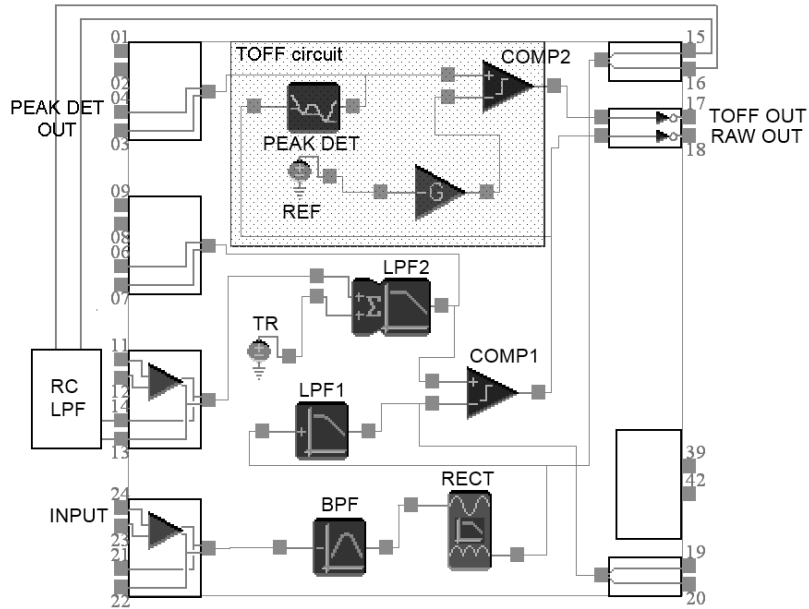}
\caption{FPAA implementation of the meteor scatter detector: INPUT - input audio signal, BPF - 10-600Hz band-pass filter, RECT – full-wave rectifier with a low-pass filter (envelope detector), LPF1 - low-pass filter 1 (20Hz), LPF2+RCLPF - low-pass filter 2 (0.1Hz), TR - threshold level adjustment,  COMP1 - comparator 1, PEAK DET - peak detector, REF - detection dead time reference level adjustment, COMP2 - comparator 2, RAW OUT - output of the detector without time off-delay circuit, TOFF OUT - output of the detector with time off-delay circuit.}
\label{f6}
\end{figure}

Our signal chain consists of several blocks (see Fig.\ref{f6}): primary band-pass filter 10Hz - 600Hz (BPF), full-wave rectifier (RECT), secondary filters (LPF1, LPF2+RCLPF), primary comparator (COMP1) and analogue realization of the time off-delay function (TOFF). We used nearly all available analogue cells of ANADIGM FPAA.

\begin{figure}[H]
\centering
\epsscale{0.8}
\plotone{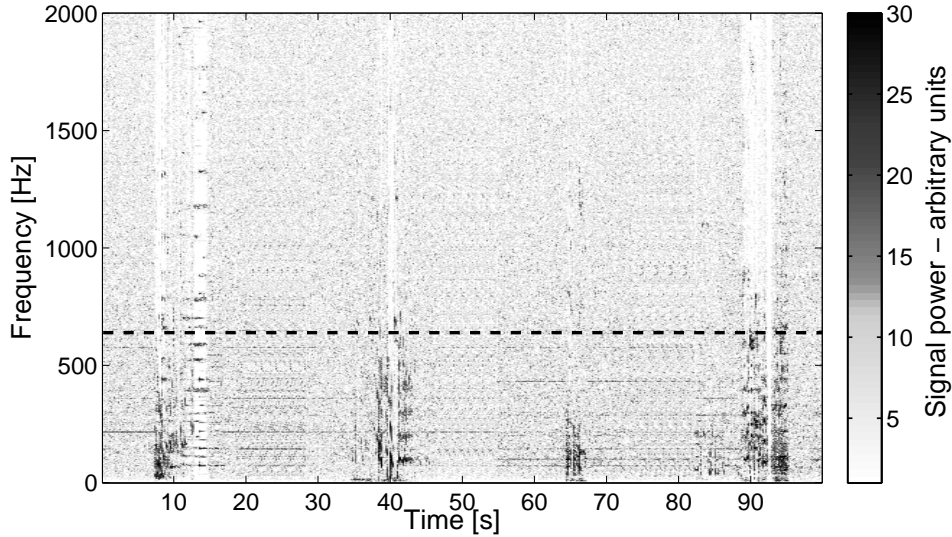}
\caption{Audio signals of four meteor scatter phenomena in the time-frequency domain. The examples include various amplitudes and different durations. By dashed line we indicate the cut-off frequency (600Hz) of our detector.}
\label{f7}
\end{figure}

As the analyzed signal is in the audio frequency range, we decided to choose a primary band-pass filter 10 - 600 Hz. It is due to the fact that most of the spectral components are within this range. As a proof, we present a plot of a part of registered radio signals in the time-frequency domain in Fig.~\ref{f7}. The corresponding audio signals are the representative examples obtained by the PFN radio meteor station in our experiment. We depicted various phenomena, with long and short duration, high and low amplitudes. We checked that all investigated audio signals in our recording were well within the assumed frequency range.

The secondary filters (LPF2 with RC LPF) provide information about the mean level of the signal power. We constantly compare the output of 20 Hz filter LPF1 with the biased (TR - threshold) output of the 0.1 Hz filter LPF2. The biasing prevents the false detections induced by temporal noise fluctuations. Both filters outputs represent the level of signal power: the first one detects rapid level changes, the second one shows a long-term trend. The first comparator (COMP1) is set high when the LPF1 output signal exceeds the LPF2 output. COMP1 output is our first-stage detector (RAW OUT).

However, we observed that during a single radio station signal appearance the RAW OUT output had several positive detections, although there should be only one meteor phenomenon detected. A solution to this problem is provided by an analogue version of a time off-delay (TOFF) circuit. It is realized with a filtered peak detector (PEAK DET), a reference signal and a comparator. When the positive edge of the first comparator (COMP1) appears, the output of the peak detector (PEAK DET) and the output of the whole device (TOFF OUT) immediately rise and are held as long as the output of the first comparator (COMP1) remains high. When it switches to the low state, the output of the peak detector (PEAK DET) slowly falls down as an RC circuit in the peak detector discharges. As long as the reference (REF) is not reached, the second comparator (COMP2) remains high. When the reference level is reached, the second comparator (COMP2) changes the output state to low, but if there is another detection in the meantime, the device output is not affected and remains high. Adjusting the reference (REF) or the RC time constant in the peak detector can change the time delay. We decided to choose a 3 second interval, which is also our estimation of the minimal interval between the subsequent meteors. The disadvantage of this method - the time delay of the detection end - can be easily overcome by reducing measured length of the phenomena by a corresponding and constant time interval. We would like to note that the detector will produce an additional detection if there is a longer than 3-second silence in the radio program. It may be easily mitigated by extending the delay; however, it also reduces the detector ability to distinguish close events.

\begin{figure}[H]
\centering
\epsscale{1.05}
\plottwo{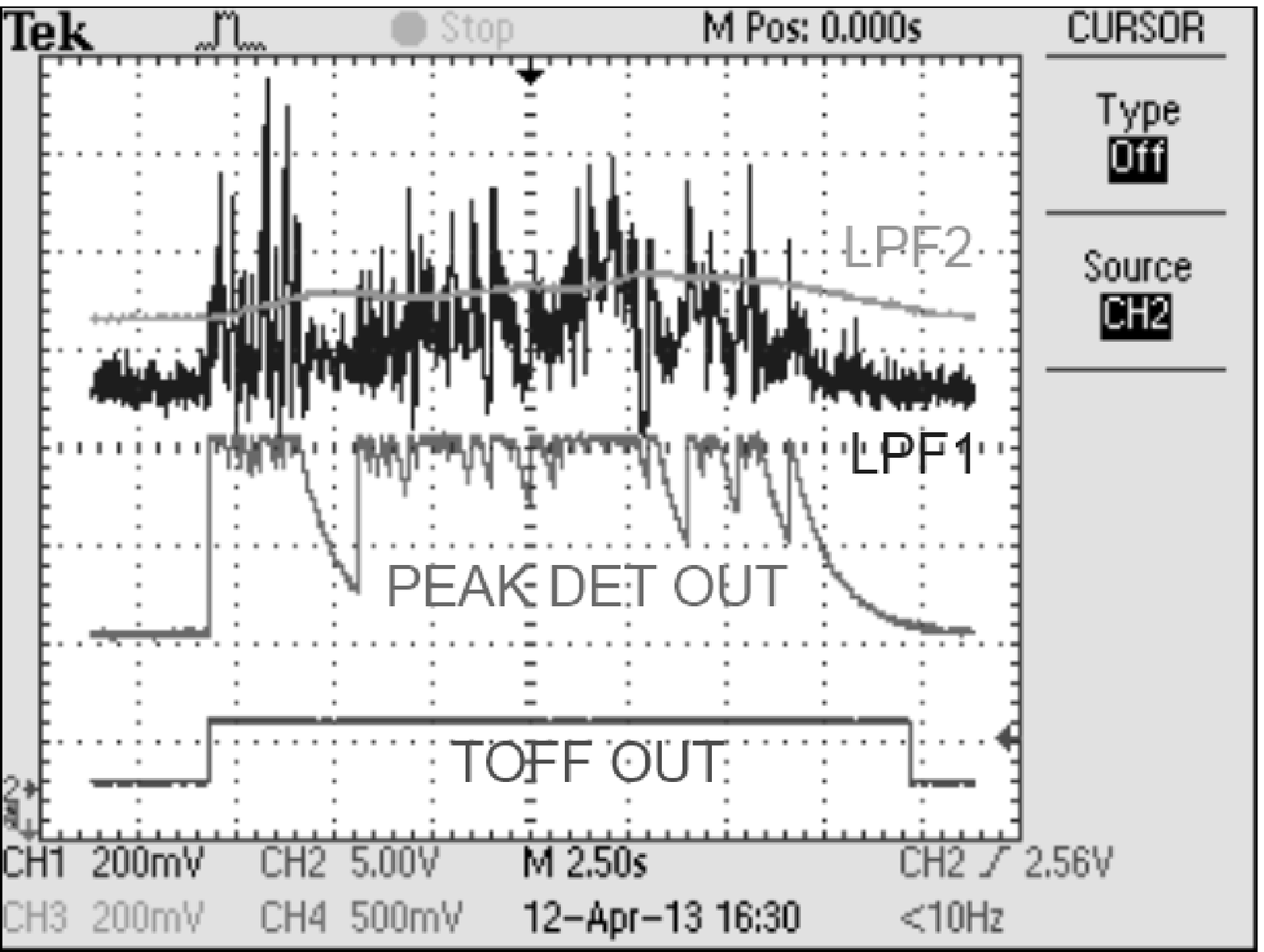}{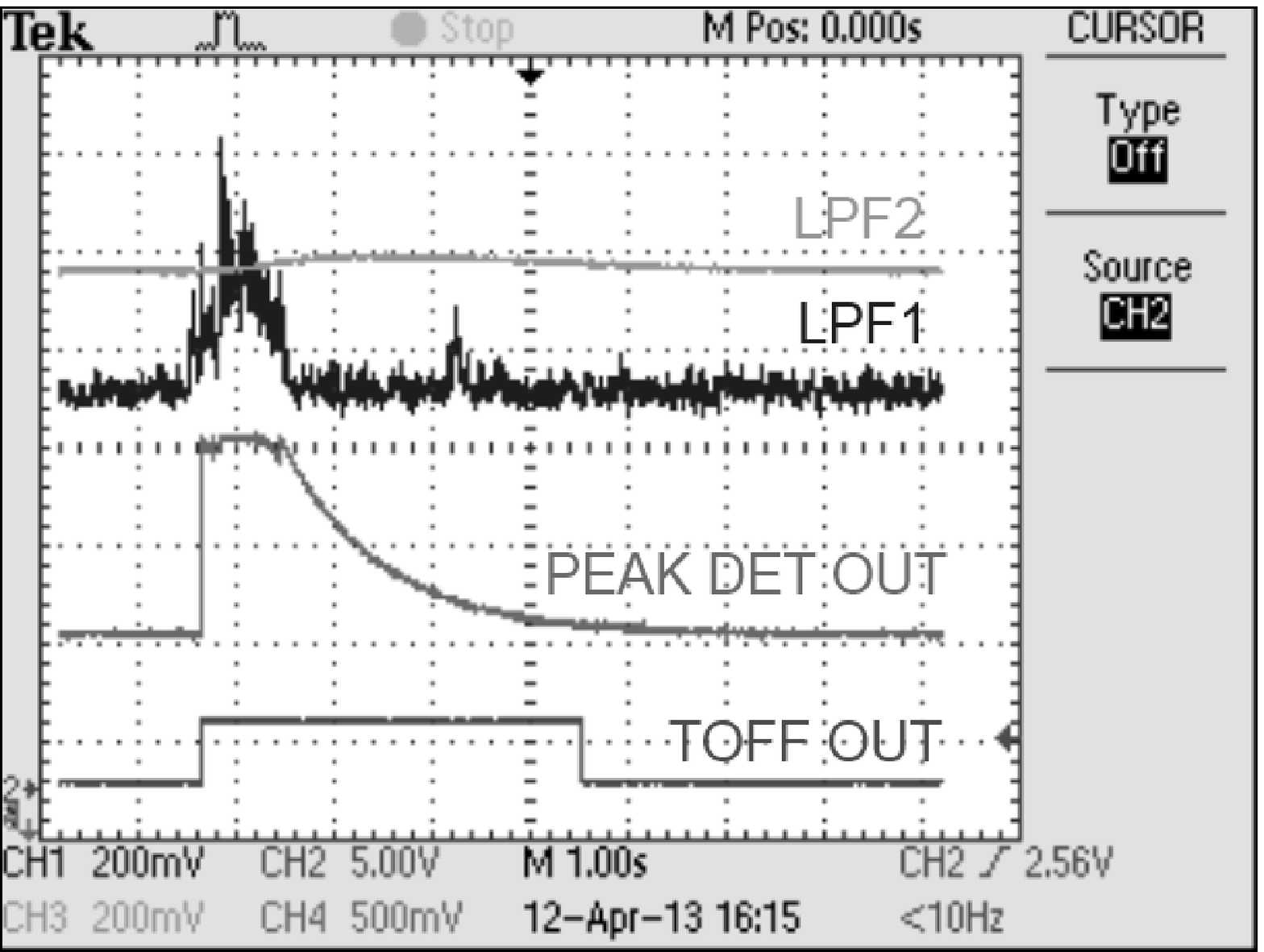}
\caption{An example of signals generated inside FPAA during a very long (over 15 seconds, left figure) and a short (several hundreds milliseconds, right figure) meteor scatter signal.}
\label{f89}
\end{figure}

The principle of operation is presented in Fig. \ref{f89}. We used two meteor scatter audio signals - a long one (on the left) and a short one (on the right) - and showed the most important waveforms obtained inside the designed FPAA circuit. The figures present the screenshots registered by the TECTRONIX TDS2014B digital oscilloscope. The detections start when the signal from LPF1 output exceeds LPF2 output. The operation of the analogue time off-delay circuit is clearly visible on the PEAK DET OUT and the TOFF OUT waveforms.

\section{Verification methodology}

\begin{figure}[H]
\centering
\epsscale{0.6}
\plotone{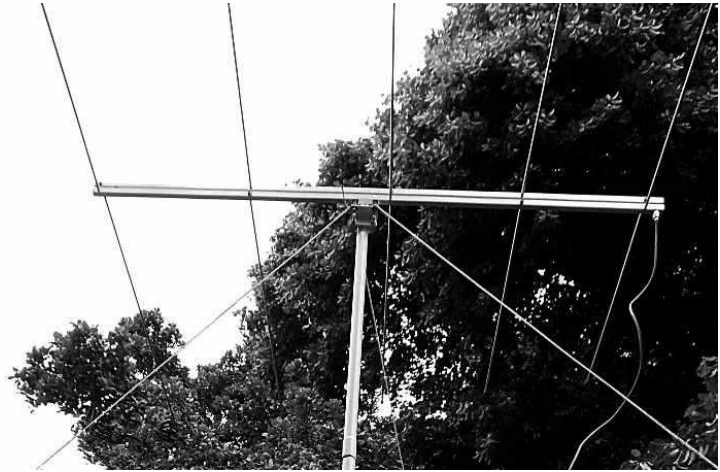}
\caption{5-element log-periodic antenna used by the Polish Fireball Network (direction: north-east, azimuth 67$^o$, horizontal polarization, height 3m). The picture was taken from below the antenna mast.}
\label{f10}
\end{figure}

We used two-hour recording provided by the PFN. Their equipment for our tests included a 5-element log-periodic antenna (Fig. \ref{f10}), 9dBi gain, provided by VPA-SYSTEM. The antenna was directed toward a distant Belorussian transmitter. We provide all necessary transmitter's information in Tab. \ref{tab1}. An audio signal from the Realistic Pro and DX-394 receiver tuned to 70.10 MHz was captured and saved on a PC computer as a \emph{wav} audio file. Since the Polish broadcast FM radio stations use the frequency band of 88-108 MHz, none of them should affect the measuring meteor scatter audio signals and cause false detections.

\begin{table}[H]
\centering
\caption{Transmitter characteristics.\label{tab1}}
\begin{tabular}{p{5cm}|p{5cm}}
\tableline
location & Mahilyow/RTPS Polykovici \\
\tableline
longitude & 30$^o$ 19' 36.41'' \\
\tableline
latitude & 53$^o$ 59' 25.92'' \\
\tableline
frequency & 70.10 Mhz\\
\tableline
program & BR Radio Mahilyow\\
\tableline
power & 4.00 kW\\
\tableline
azimuth from receiver & 67$^o$\\
distance from receiver & 667 km\\

\tableline
\end{tabular}
\end{table}

As the detection algorithms implemented in the MetAn software and in our detector are highly dependent on a threshold level, we decided to repeat the detection procedure for several values of the threshold. We did it for both the MetAn software and our hardware detector. For this purpose we used several parts of previously mentioned recording, to reduce the measurement time for each sensitivity level. The constructed, well defined, test signal included 40 clearly audible signals extracted from 2-hour recording. They were combined with the proceeding and following 10 seconds, thus our test recording was about 15 minutes long (instead of 2 hours).

\begin{figure}[H]
\centering
\epsscale{1.1}
\plottwo{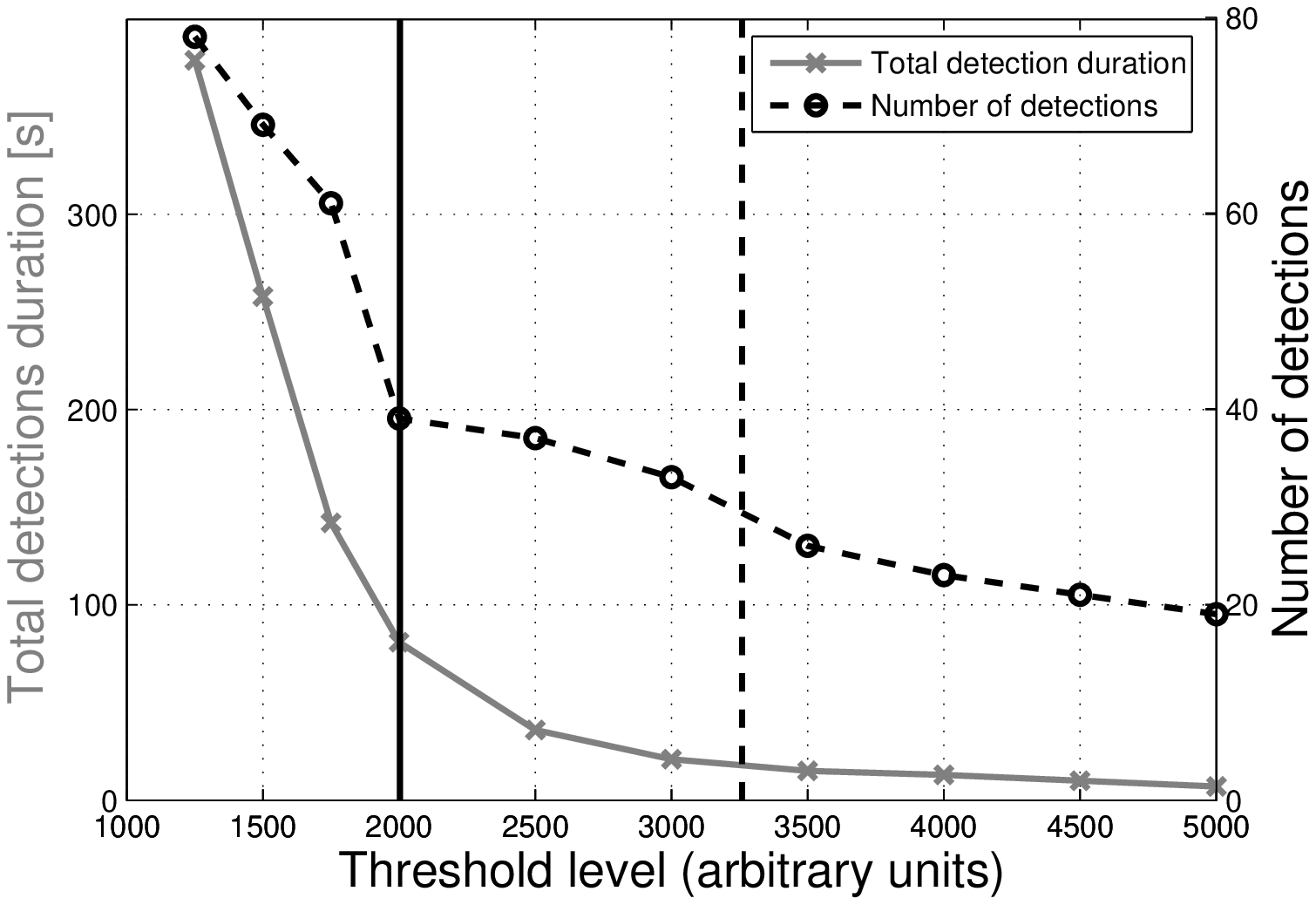}{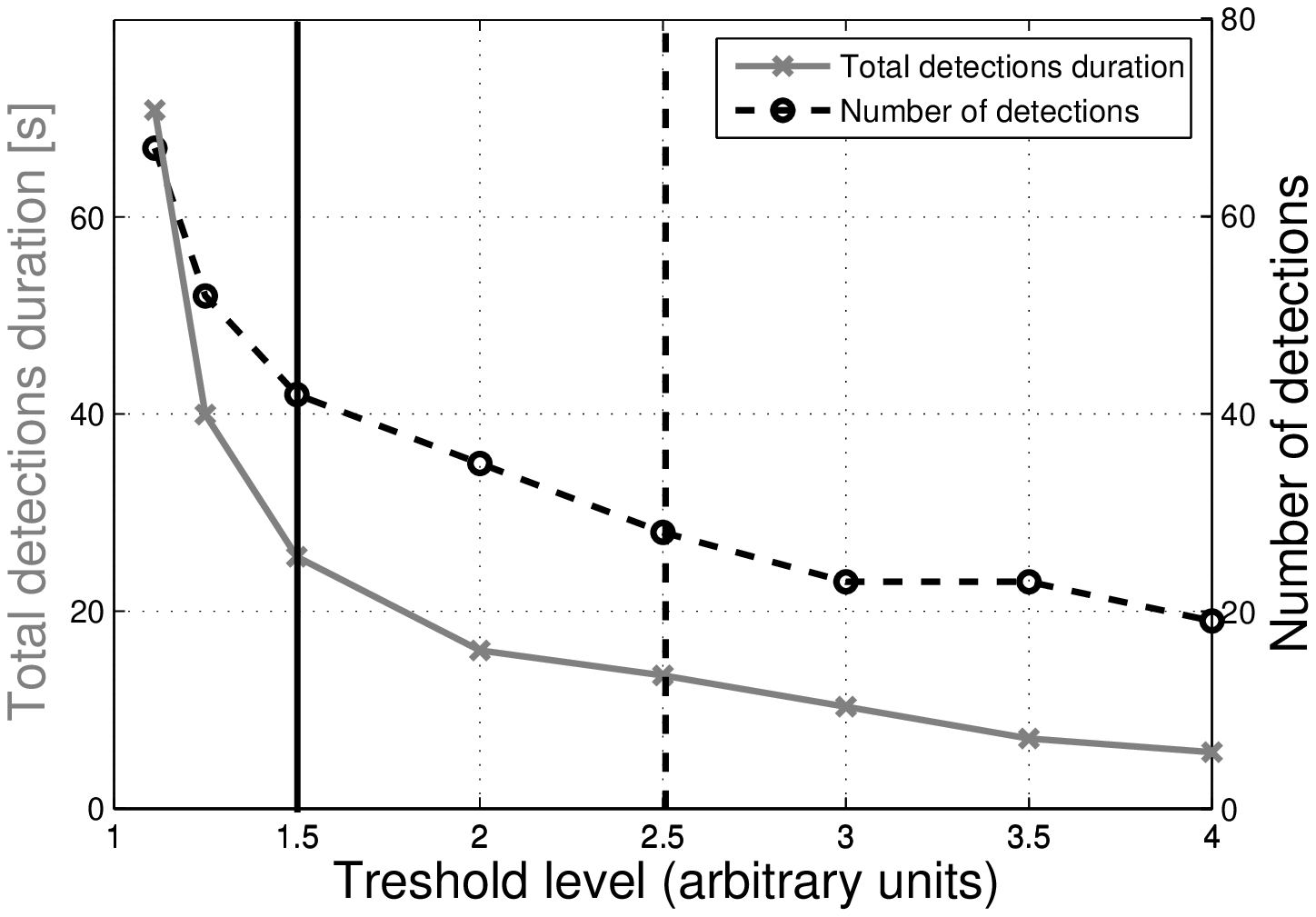}
\caption{Number of detections (dashed black line) and total detections duration (solid gray line) as a function of the threshold level for the MetAn software (on the left) and for our FPAA based detector (on the right). By the vertical solid line we suggest the threshold level for maximum sensitivity. By the vertical dashed line we suggest the threshold level for medium sensitivity.}
\label{f1112}
\end{figure}

The resulting dependencies of the number of detections and the total detections duration are presented in Fig. \ref{f1112}. It is visible that each resulting curve shows a characteristic break related to the rising number of false detections due to the noise level comparable to the threshold. We decided to set both detectors to their maximum and medium sensitivities, which are represented by the vertical solid and dashed lines in both corresponding figures. 

Firstly, we selected the maximum sensitivity thresholds, for which the number of obtained detections was as close as possible to the number of given phenomena (40). It also let us estimate the thresholds that should not be exceeded to avoid a rapidly rising number of false detections. The medium sensitivity thresholds were chosen for the number of detections equal to about 30. It corresponds to the situation, where the detectors have a slightly higher detection threshold and consequently they miss about 10 weakest signals. 

We would like to note, that the total detections duration for MetAn software is significantly larger because we did not know the additional time interval added after each detection. In the case of our hardware detector the delay of detection end was known - exactly 3 seconds - thus we were able to remove this additional offset.

Finally, the main recording (2 hours long) was given at the FPAA input and at the line-in input of a PC computer which was running the MetAn software detector. We observed simultaneously the output of both detectors and the input signal. The detections for both medium and  maximum sensitivities were compared. All events detected by both detectors were carefully examined using headphones to check if the audio signal was audible. In this way we classified detections as the false and the true ones.

\section{Results}

\begin{figure}[H]
\centering
\epsscale{1.4}
\plottwo{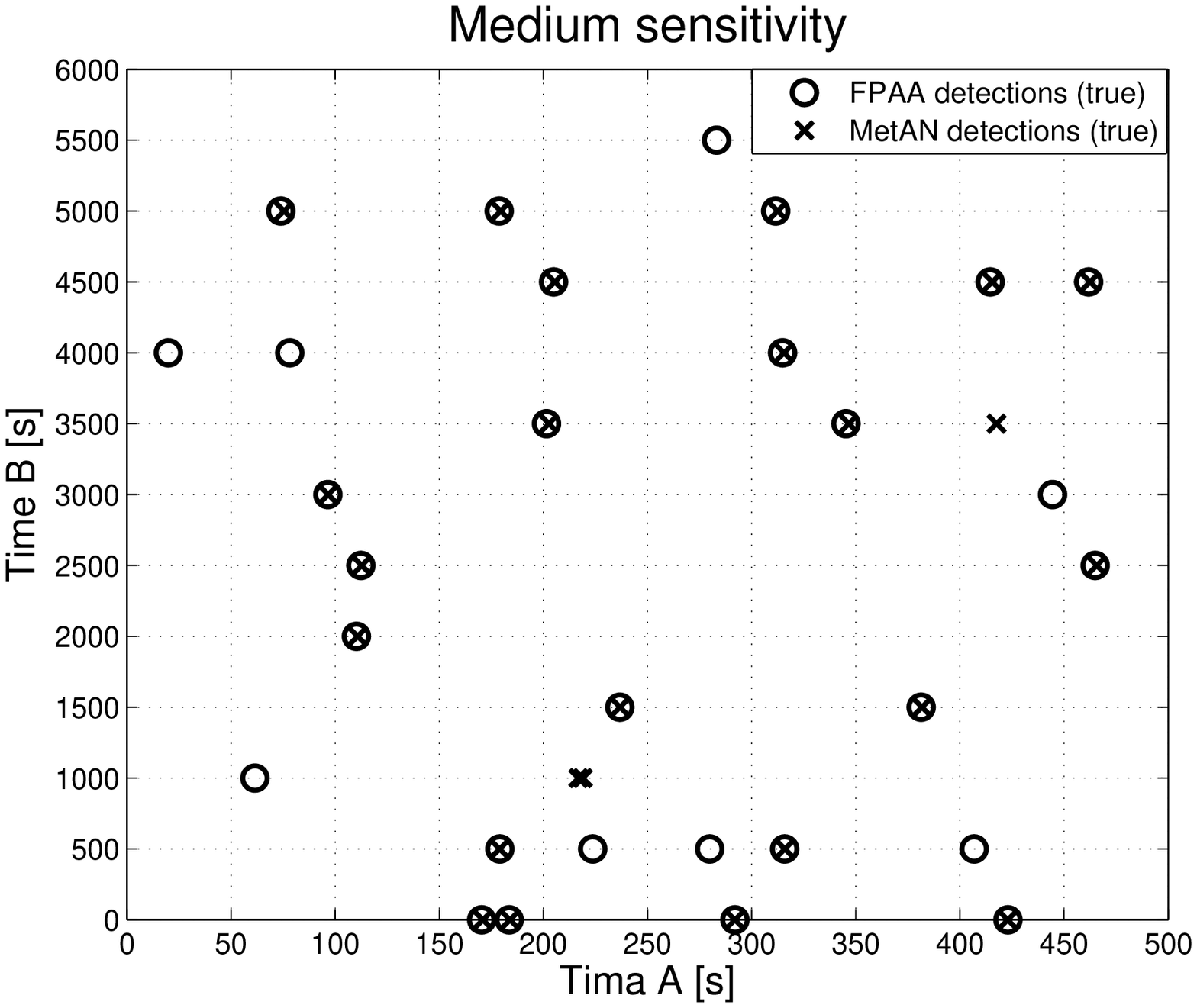}{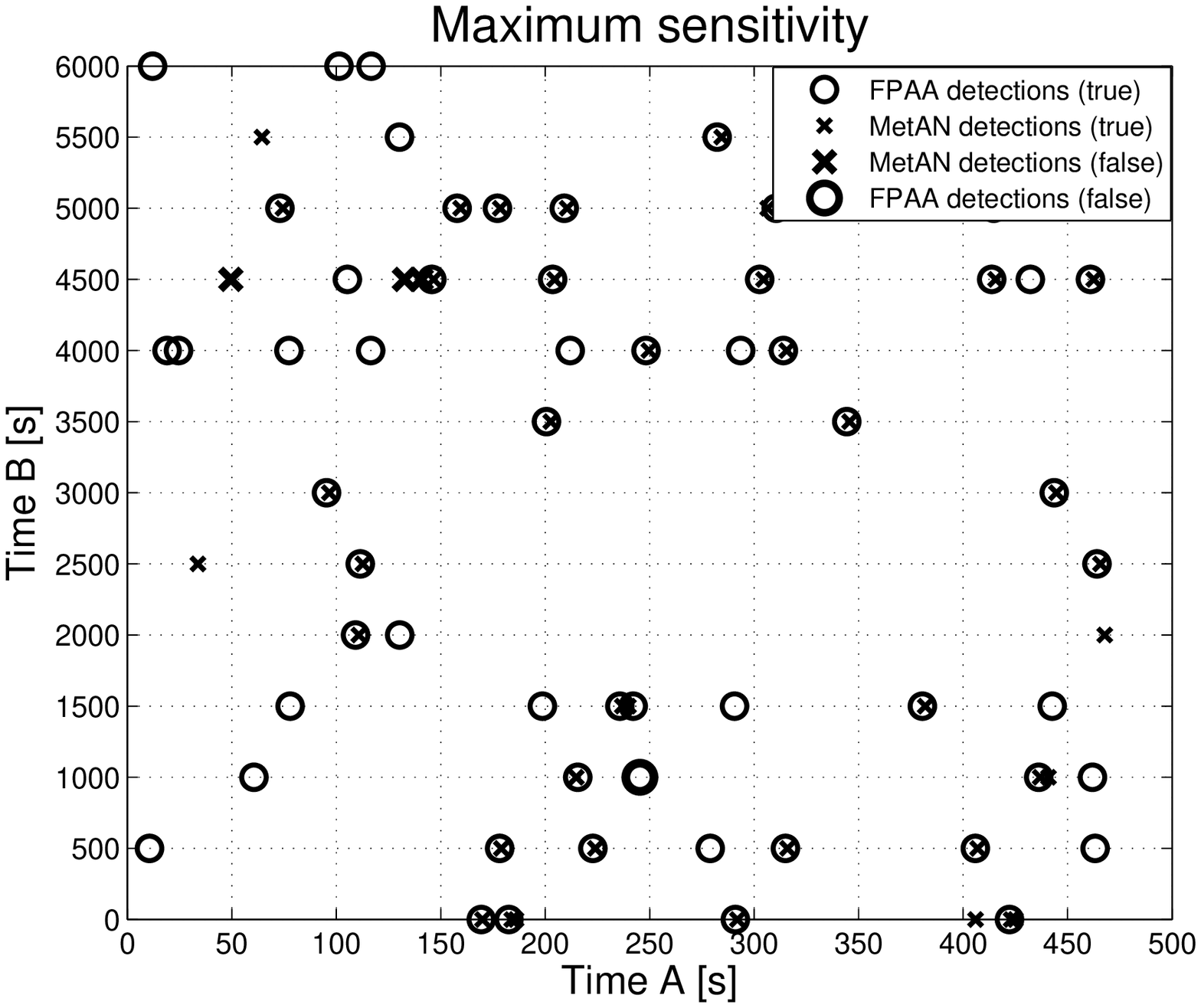}
\caption{Time plot of detections for medium (upper plot) and maximum (lower plot) sensitivities. The event time can be calculated by adding TimeA and TimeB.}
\label{f1314}
\end{figure}

The results are presented in Fig. \ref{f1314}. Each event time can be computed by adding TimeA and TimeB. The total duration of our recording (2 hours) was reduced to 6500 seconds in Fig. \ref{f1314} as the last part did not include any detection. We chose full circles and squares to identify false detections of respectively FPAA and MetAN detector, empty circle markers for positive FPAA detections and crosses for MetAN positive detections.

The results for both sensitivities show that many of the detections overlap. Almost all detections of the MetAN software were also detected by FPAA (except for 4 and 2 detections respectively for maximum and medium sensitivities), which confirms that FPAA covers the domain of MetAN. However, we would like to indicate that our detector found a lot of additional events in the full recording, which were not recognized by the software detector (red circles without blue crosses inside, respectively 22 for maximum and 8 for medium sensitivity).  Moreover, the number of false positive detections is higher for the MetAN analyzer, thus our detector is more resistant to noise fluctuations, which may influence the detector's decision. We summarize the results in Table~\ref{tab2}.

The quality of measurement, hence the quality of obtained results, depends on both the number of true detections (the signal) and the number of false detections (the noise floor). Ideally, the detector should maximize the signal level while keeping the low noise floor (i.e. maximize the signal-to-noise ratio). The results of our experiment for both sensitivities prove that this done much better in our solution than in the competitive detector.

\begin{table}[H]
\centering
\caption{Detections summary for both detectors and for both sensitivities.\label{tab2}}
\begin{tabular}{p{2.7cm}|p{1.5cm}p{1.5cm}p{1.5cm}|p{1.5cm}p{1.5cm}p{1.5cm}}
\tableline

&\multicolumn{3}{c|}{Medium sensitivity}&\multicolumn{3}{c}{Maximum sensitivity} \\

&FPAA&MetAN&Common&FPAA&MetAN&Common\\
\tableline  
True positive&29&23&21&55&37&33\\ \tableline
False positive&0&0&0&1&3&0\\ \tableline
\end{tabular}

\end{table}

\section{Conclusions}

In the article we presented a hardware meteor radio detector which can be attached to any radio device to detect the meteor scatter phenomena. The signal processing circuit of our detector was implemented in ANADIGM field programmable analog array (FPAA) using nearly all its available resources.  A signal path contains a rectifier, a peak detector, several comparators, low-pass and band-pass filters. The output signals of the filter provide information about signal power in the audio frequency range. Additionally, the implemented time off-delay analogue circuit prevents multiple detections during long scatter phenomena. Such an FPAA device enabled flexible signal flow designing to fulfill the detector requirements.

We compared the efficiency of detection of our implementation with the current solution used by the Polish Fireball Network (PFN) -  the MetAN software dedicated to a PC computer. We presented the results of the analysis of 2-hour recording received from a PFN meteor station receiving Belarusian radio signals. By setting both detectors to medium and maximum sensitivities we compared the effectiveness of both solutions for different detection threshold levels.

The results proved a significant difference between the detectors. The FPAA detector proved its ability to pick up noticeably more audio signals for both thresholds. When set to the maximum sensitivity, it detected even hardly audible scatter events while maintaining its low false detection rate (only 1 false positive per 57 detections). In comparison, current software detector found significantly fewer events and had lower true detection percentage. Our experiment confirmed that the measurements employing FPAA detector are of higher efficiency, thus the outcomes have a better signal to noise ratio.

Currently, we are developing a ready-to-go electronic board with peripheries incorporating the presented FPAA-based detector, recording on an SD-card and an optional screen presenting counts histograms. Such a device could be easily handled not only by professional meteor researchers, but also by amateur astronomers who wish to contribute to additional radio observations. With its efficiency and very little power consumption (comparing to a PC computer), our detector might replace current software-based detecting methods. We also wish to set a network of remote detectors which will provide complex data and enable further investigation of its radiolocation capabilities.

\acknowledgments

ACKNOWLEDGMENTS: Adam Popowicz was financed by NCBiR project no POIG.02.03.01-24-099/13 “Upper Silesian Center for Computational Science and Engineering”. 

Andrzej Malcher and Krzysztof Bernacki were financed by Principal Research Grants BK of the Institute of Electronics at the Silesian University of Technology.

The research was performed using the infrastructure supported by POIG.02.03.01-24-099/13 grant: GCONiI - Upper-Silesian Center for Scientific Computation.

We are grateful to the Polish Fireball Network team (PFN), who shared with us their specialist radio equipment used during the signal acquisition. We would like to thank anonymous referee for all helpful and constructive comments and suggestions.

\end{document}